\documentclass[a4paper]{spie}
\usepackage[]{graphicx}
\usepackage[latin1]{inputenc}
\usepackage{array}
\usepackage{amsmath,amssymb}
\usepackage{gensymb}
\usepackage{color}
\usepackage{mathrsfs}
\usepackage{stmaryrd}
\DeclareTextSymbol{\degre}{OT1}{23}
\def\registered{\ooalign{\hfil\raise .00ex\hbox{\scriptsize R}\hfil\crcr\mathhexbox20D}}

\title{A Mach-Zehnder interferometer based on orbital angular momentum for improved vortex coronagraph efficiency} 
\author{Piron P.\supit{a}, Delacroix C.\supit{b}, Huby E.\supit{a}, Mawet D.\supit{c}, Karlsson M.\supit{d}, Ruane G.\supit{e}, Habraken S.\supit{a}, Absil O.\supit{a} and Surdej J.\supit{a}
\skiplinehalf
\supit{a} Department of Astrophysics, Geophysics and Oceanography, University of Liège, 17 Allée du 6 Août, 4000 Liège, Belgium;\\%
\supit{b} High Angular Resolution $\&$ Stellar Surroundings Astrophysics (HARISSA), Centre de Recherche de LYON (CRAL), 9 Avenue Char les André, F -69561 Saint-Genis-Laval, France;\\%
\supit{c} Astronomy Department, California Institue of Technology,1200 E. California Blvd., Pasadena California 91125, USA;\\%
\supit{d} \AA ngström Laboratory, Uppsala University Lägerhyddsvägen 1, SE-751 21 Uppsala, Sweden;\\%
\supit{e} Chester F. Carlson Center for Imaging Science, Rochester Institue of Technology, 54 Lomb Mem. Dr., Rochester NY 14623, USA;
}
\authorinfo{Further author information: (Send correspondence to Piron P.)\\Piron P.: E-mail: pierre.piron@ulg.ac.be}
   \begin{document} 
  \maketitle 
\begin{abstract} 
The Annular Groove Phase Mask (AGPM) is a vectorial vortex phase mask. It acts as a half-wave plate with a radial fast axis orientation operating in the mid infrared domain. When placed at the the focus of a telescope element provides a continuous helical phase ramp for an on axis sources, which creates the orbital angular momentum. Thanks to that phase, the intensity of the central source is canceled by a down-stream pupil stop, while the off axis sources are not affected. However due to experimental conditions the nulling is hardly perfect. 
To improve the null, a Mach-Zehnder interferometer containing Dove prisms differently oriented can be proposed to sort out light based on its orbital angular momentum (OAM). Thanks to the differential rotation of the beam, a $\pi$ phase shift is achieved for the on axis light affected by a non zero OAM. Therefore the contrast between the star and its faint companion is enhanced. Nevertheless, due the Dove prisms birefringence, the performance of the interferometer is relatively poor. To solve this problem, we propose to add a birefringent wave-plate in each arm to compensate this birefringence.\\%
In this paper, we will develop the mathematical model of the wave front using the Jones formalism. The performance of the interferometer is at first computed for the simple version without the birefringent plate. Then the effect of the birefringent plate is be mathematically described and the performance is re-computed.
\end{abstract}

\keywords{Orbital angular momentum, Coronagraphy, Interferometry, Birefringence,Polarization}

\section{INTRODUCTION}
\label{sec:intro} 
\begin{figure}[h!]
\begin{center}
\includegraphics[width=0.6\columnwidth]{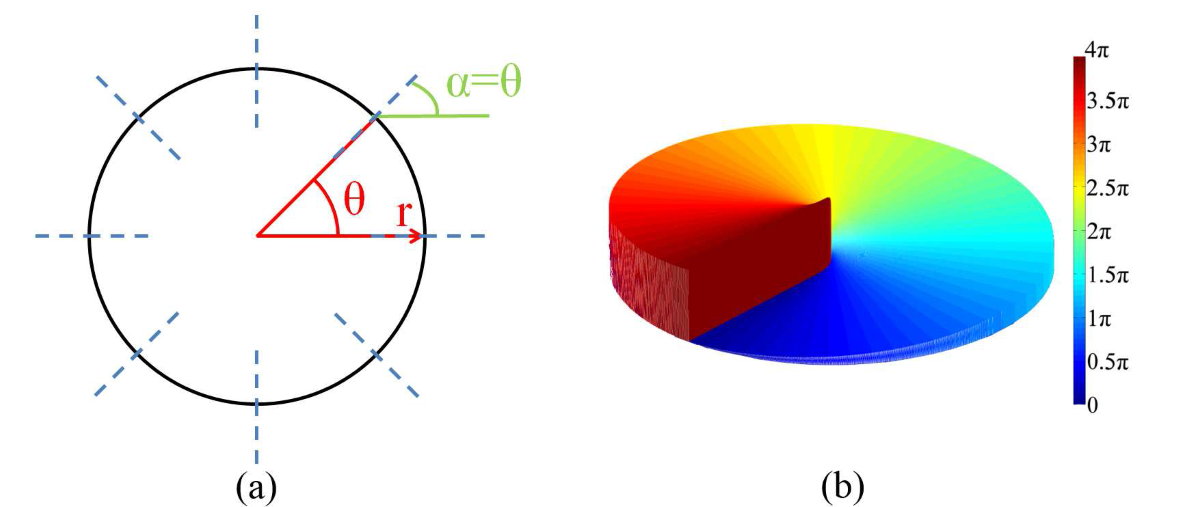}%
\caption{\label{fig:AGPM} (a) Representation of the AGPM where the fast axis orientation is represented by dashed blue  lines; (b) representation of the phase distribution.}%
\end{center}
\end{figure}

\begin{figure}[h!]
\begin{center}
\includegraphics[width=0.7\columnwidth]{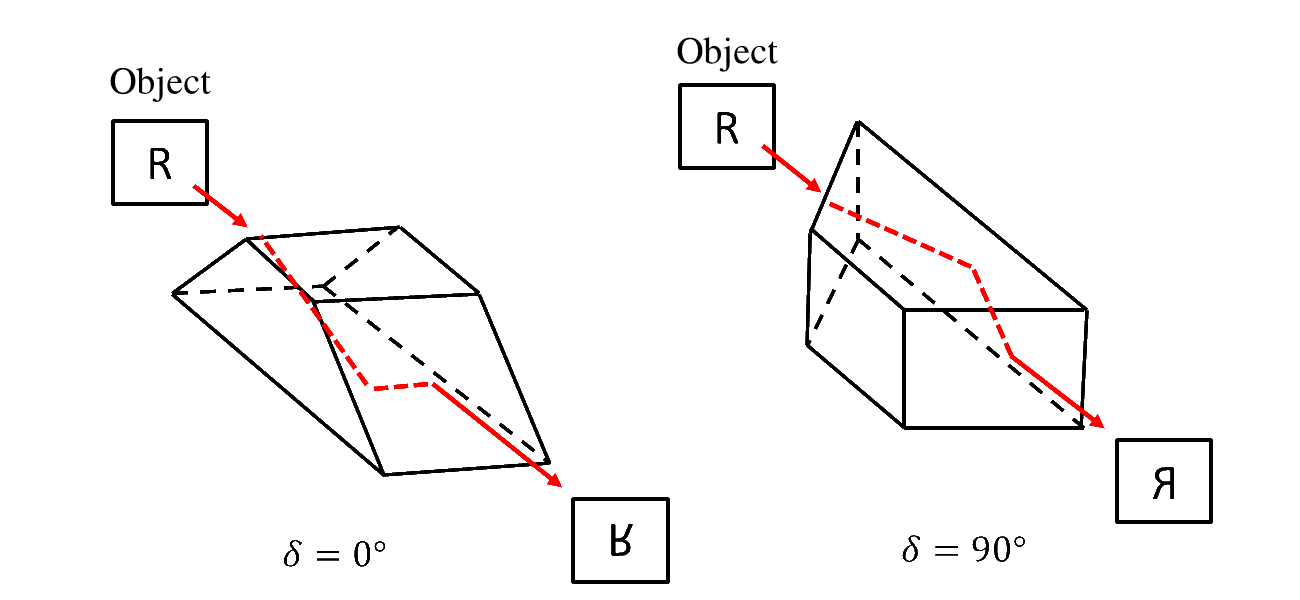}%
\caption{\label{fig:Dov} Picture of the Dove prisms and its effect on the image for two different orientations. For the right-hand side prism the effect can be viewed as the superimposition of the image flip of the left prism with a rotation of 180\degre.}%
\end{center}
\end{figure}

\begin{figure}[h!]
\begin{center}
\includegraphics[width=0.5\columnwidth]{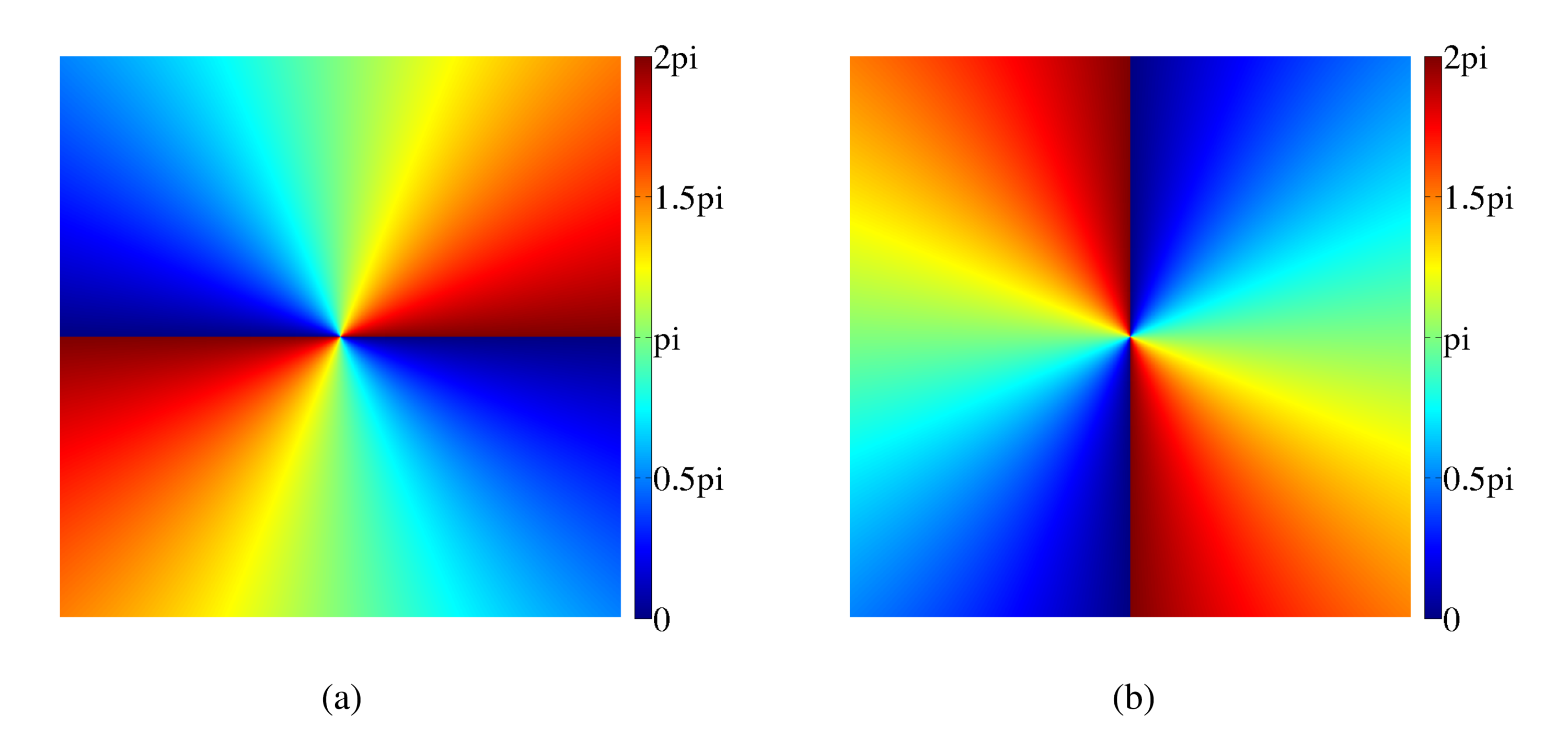}%
\caption{\label{fig:pha} Representation of the phase distribution for an on-axis source for the beams in the two interferometer arms. The difference between the two phase distributions is a rotation of 90\degree due to the Dove prisms. A $\pi$ phase difference is present between the two beams at beam combination and destructive interference occurs.}%
\end{center}
\end{figure}

The goal of phase mask coronagraphy is to reduce the light of a central star to improve the contrast of its faint companions. The Vortex team in Liège and Uppsala already developed, manufactured and installed several phase masks~\cite{Maw10,Maw13,Del13,Def14,Maw14}. The goal of our system is to improve the performances of the Annular Groove phase Mask (AGPM) on a coronagraphic configuration. The AGPM which is composed of subwavelength gratings arranged in a circular pattern. It acts as a half-wave plate with a radial orientation of its fast axis (see Figure~\ref{fig:AGPM} (a)). After this component, the transmitted beam is characterized by a phase dislocation of the form $\exp\left(il\theta\right)$ (see Fig~\ref{fig:AGPM} (b)) where $l$ is the topological charge\cite{Niv06,Gal02}, with $l=2$ for the AGPM, and $\theta$ is the radial angle.\\%
\newpage
However, due to experimental conditions (deviation from the half-wave plate, presence of a central obstruction on the primary mirror, atmospheric turbulence, etc), the attenuation of the central starlight is not perfect and a residual light remains. 
It has been demonstrated that beams with a helical phase $\Phi=l\theta$ carries an Orbital Angular Momentum  (OAM) $l\bar{h}$ \cite{Yao11,All92}. Inspired by methods to determine the OAM \cite{Lea02,Gao10,Lav11}, we plan to use the specific phase pattern produced by the AGPM and an interferometer based on Dove prisms to improve the rejection at the detector as proposed by Riaud et al .\\%
The goal of the Dove prisms is to induce a rotation of the incident electric field (see Figure~\ref{fig:Dov}).\\%
Thanks to a different orientation between the prisms and the helical phase profile of the beam, a $\pi$ phase shift can be created between the two arms of the interferometer for the central source (see Figure \ref{fig:pha}) without causing destructive interference for the off axis sources. Therefore, the contrast ratio between the off axis sources and the residual central starlight is improved increasing the detection probability for off axis sources.\\%
This setup is a simplified version of the one proposed by P. Riaud~\cite{Ria14} where four Dove prisms are implied to correct the differential rotation and overcome possible wavefront aberrations. 

In this paper, we focus on the effect of the Dove on the polarization and their impact on the interferometer performance. Firstly, the optical setup and its mathematical model will be described. Next the simulation results will be presented and analyzed. Then, an improved version containing additional birefringent plates will be proposed. Finally, the main results will be summarized and future work will be discussed.
\newpage
\section{OPTICAL SETUP}
\label{sec:exp}
\begin{figure}[h!]
\begin{center}
\includegraphics[width=0.47\columnwidth]{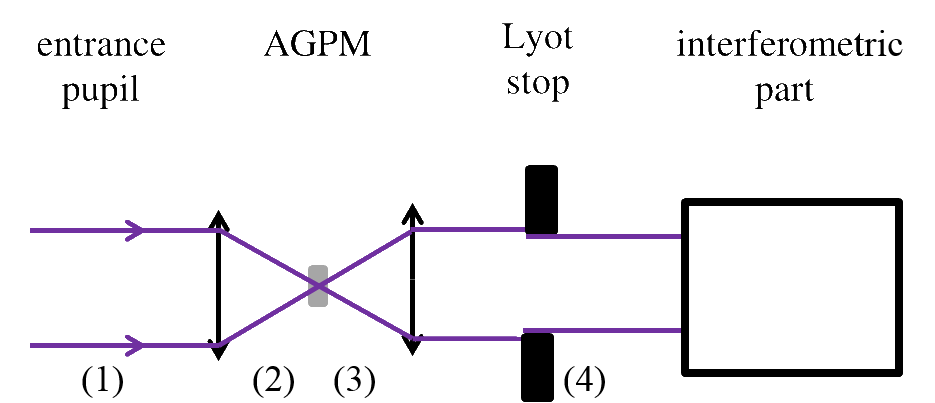}%
\caption{\label{fig:corpart} Representation of the coronagraphic part of the optical setup.}%
\end{center}
\end{figure}

\begin{figure}[h!]
\begin{center}
\includegraphics[width=0.8\columnwidth]{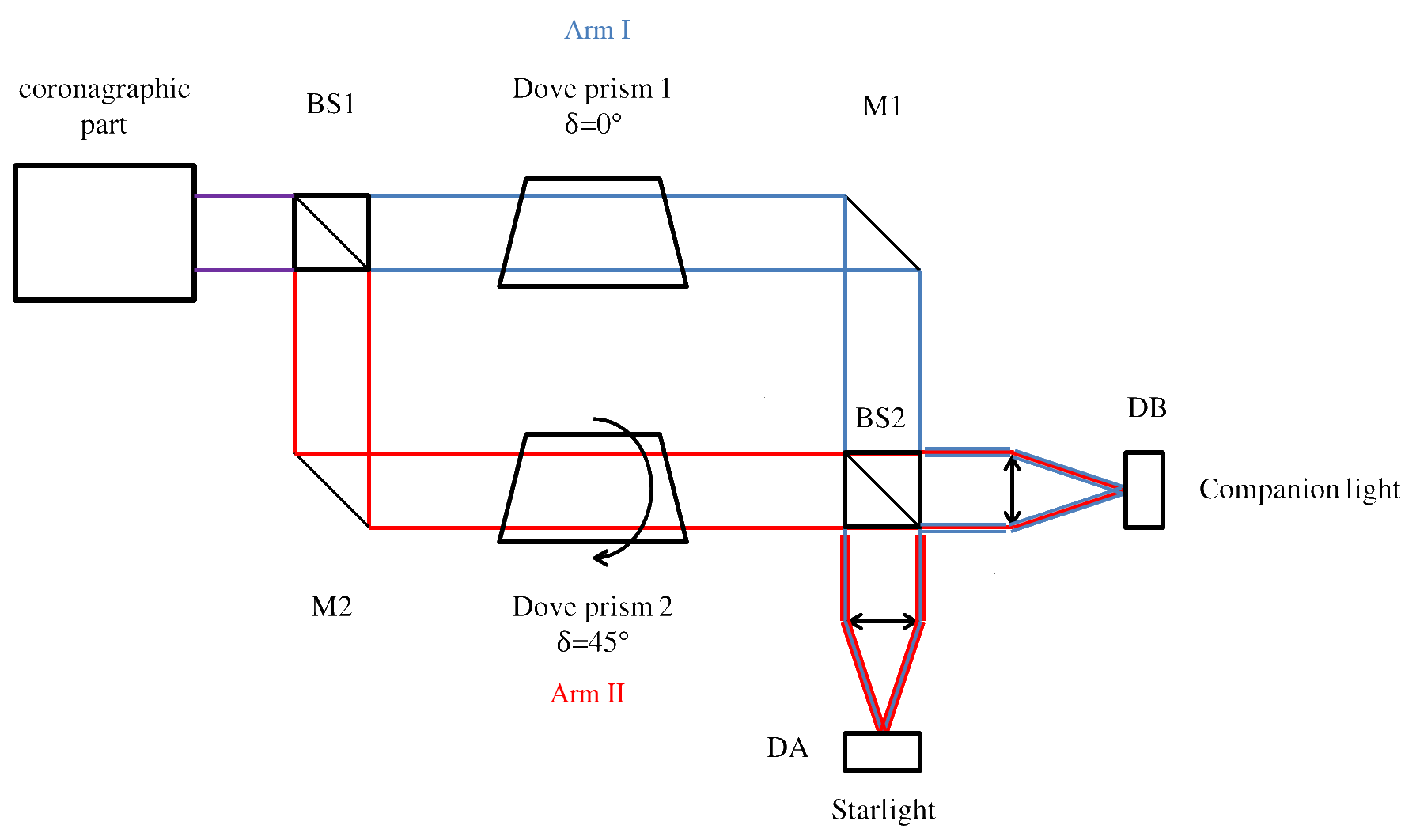}%
\caption{\label{fig:interfpart} Representation of the interferometric part of the optical setup.}%
\end{center}
\end{figure}

The experimental setup is depicted in Figures~\ref{fig:corpart} \& \ref{fig:interfpart}.
The setup can be divided into two parts. The first one concerns the generic coronagraph setup while the second one contains the interferometer.

\vspace{-0.6cm}

\newpage
\subsection{Mathematical model}
The Jones formalism and the Fourier transform approach were chosen to describe and simulate our system.
At each step, the electric field will be divided into the x and y components $E=\left(\begin{array}{c}
E_{x}\\%
E_{y}\\%
\end{array}\right)$ and the Fourier transform will be applied to each component separately $FT(E)=\left(\begin{array}{c}
FT\left(E_{x}\right)\\%
FT\left(E_{y}\right)\\%
\end{array}\right)$.\\%
The AGPM and the prisms will be modeled by their Jones matrices.\\%
In our simulation, the case of a circular pupil with a $10\%$ central obstruction combined with perfect mirrors and beam-splitters without phase variation introduced by the coating was considered.
\begin{equation}
E_{Pup}=\left[\Pi\left(\dfrac{r}{2R_{Pup}}\right)-\Pi\left(\dfrac{r}{2R_{obs}}\right)\right]\times E_{in}
\end{equation}
Where $R_{pup}$ is the radius of the entrance pupil, $R_{obs}$ is the radius of the central obstruction and $E_{in}$ is the incident electric field.\\%
Before the AGPM, the field is the Fourier transform of the field at the pupil plane.
\begin{equation}
E_{preAGPM}=FT\left(E_{Pup}\right)
\end{equation}
After the AGPM, the field is multiplied by the matrix corresponding to the AGPM.
\begin{equation}
\begin{array}{c}
AGPM=\left(\begin{array} {c c}
\cos(2\theta)&\sin(2\theta)\\%
\sin(2\theta)&-\cos(2\theta)\\%
\end{array}\right)\\%
E_{postAGPM}=AGPM\times E_{preAGPM}\\%
\end{array}
\end{equation}
Where $\theta$ is the orientation of the fast axis of the birefringent plate and it corresponds to a radial orientation for the AGPM.\\%
At the Lyot stop plane, the field is the multiplication of the Fourier transform of the previous one multiplied by the Lyot stop.
\begin{equation}
E_{Lyot}=FT\left(E_{postAGPM}\right)\times \Pi\left(\dfrac{r}{2R_{Lyot}}\right)
\end{equation}
Where $R_{Lyot}$ is the radius of the Lyot stop and $R_{Lyot}<G R_{Pup}$ with $G$ the magnification. This field will be injected in the interferometer.\\%
The first element of the interferometer is a beam splitter, it will divide the incident beam into two beams. The first beam ($E_{I}$) will be transmitted and its intensity will be divided by 2. The second beam ($E_{II}$) will be reflected and have the same division of its intensity and a $\pi/2$ phase retard will be added.
\begin{equation}
\arraycolsep=1.4pt\def\arraystretch{2.2}
\begin{array}{c}
E_{I}=\dfrac{\sqrt{2}}{2}\times E_{Lyot}\\%
E_{II}=\dfrac{\sqrt{2}i}{2}\times E_{Lyot}
\end{array}
\end{equation}
Before the Dove prisms, the beam in the second part is reflected by a mirror. For a perfect mirror, a $\pi$ phase retard will occur.
\begin{equation}
E_{IIr}=-E_{II}
\label{eq:MaII}
\end{equation}
After the Dove prisms, the field is multiplied by the Jones matrix\cite{Mor04} of the Dove prisms $D_{I}$ and $D_{II}$.\\%
\begin{equation*}
\arraycolsep=1.4pt\def\arraystretch{1.5}
 D_{i}=\left(\begin{array}{c c}
-T_{//}\cos^{2}(\delta{i})-T_{\perp}\sin^{2}(\delta{i})& \left(T_{\perp}-T_{//}\right)\cos(\delta{i})\sin(\delta{i})\\%
\left(T_{\perp}-T_{//}\right)\cos(\delta{i})\sin(\delta{i}) & -T_{//}\sin^{2}(\delta{i})-T_{\perp}\cos^{2}(\delta{i})
\end{array}\right)
\end{equation*}
Where $\delta_{i}$ is the orientation angle of the Dove prisms ($\delta_{I}=0$\degre and $\delta_{II}=45$\degre) $T_{//}$ and $T_{\perp}$ are the transmission coefficients for the parallel and orthogonal components of the polarized beam. They depend on the refractive index of the prism and on the angle of the prism edges. They are described in appendix \ref{app:tran}.
\begin{equation}
\begin{array}{c}
E_{DI}=D_{I}\times E_{I}\\%
E_{DII}=D_{II}\times E_{IIB}
\label{Doveeff}
\end{array}
\end{equation}
The effect of the Dove prisms is an image flip the inversion of top and bottom for $D_{I}$ and an image followed by rotation of 90\degree for $D_{II}$ (see Figure~\ref{fig:Dov}). These are separately applied to each components of the electric field.

\begin{figure}[h!]%
\begin{center}
\includegraphics[width=0.9\columnwidth]{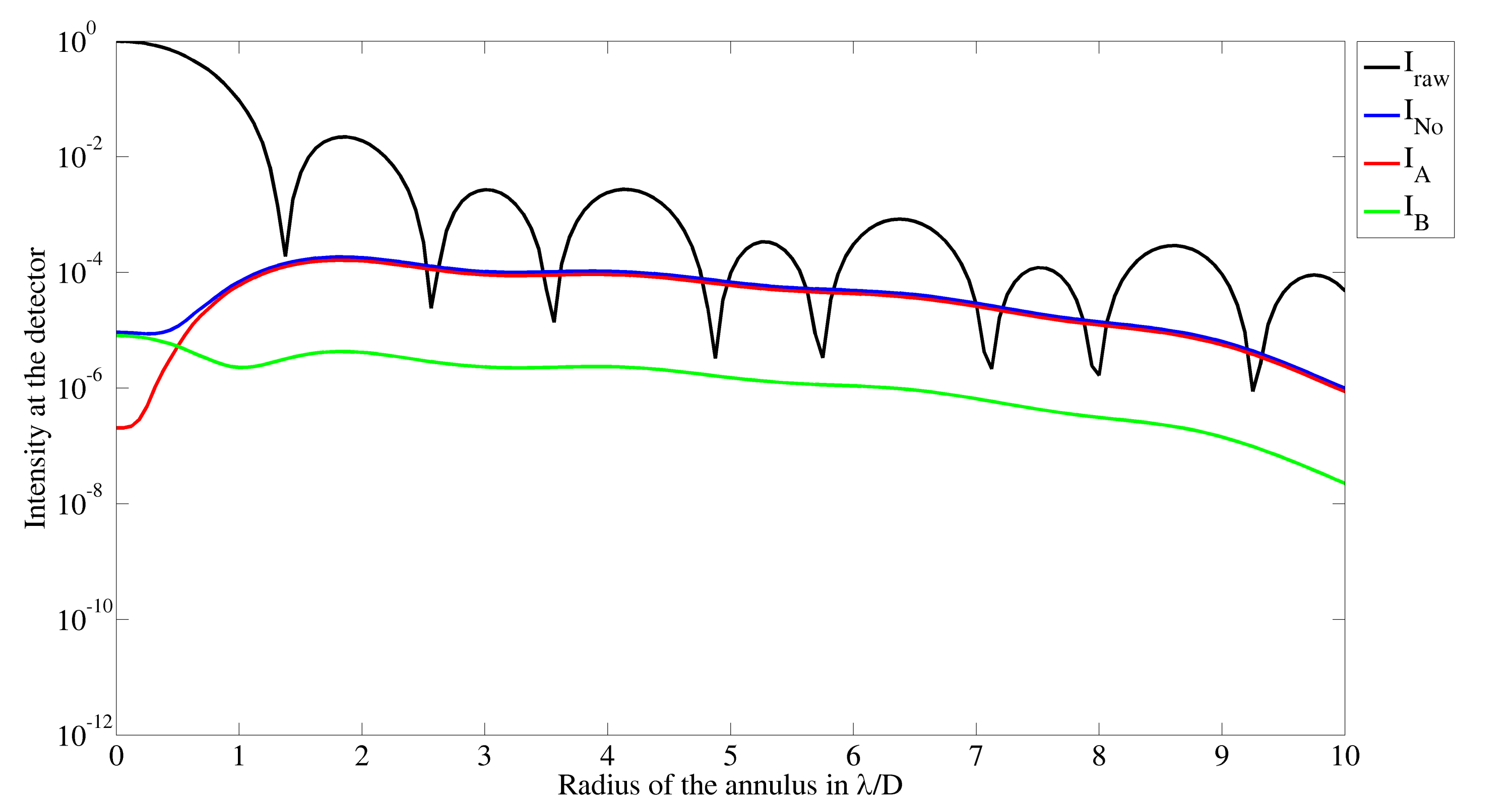}%
\caption{\label{fig:I045} Radial profile of the intensity at the different detectors for a central source with a central obstruction of 10$\%$. $I_{raw}$ represents the intensity obtained without a coronagraph and without the interferometer. The other intensities are obtained without the interferometric setup ($I_{No}$) and at the two detectors $I_{A},\,I_{B}$. The intensities are normalized with the maximum of $I_{raw}$.}%
\end{center}
\end{figure}

In the first arm before the beam splitter, the beam is reflected by a mirror.  Just like in equation \ref{eq:MaII}, a $\pi$ phase retard will occur.
\begin{equation}
E_{DIr}=-E_{DI}
\end{equation}
At the second beam splitter, each beam is divided into two other beams, which will be focused on a separate detector.\\%
For arm $I$, the beam will be transmitted into the A path and reflected into the B path while for arm $II$, the A path is the reflected one while the B path is the transmitted one.
\begin{equation}
\arraycolsep=1.4pt\def\arraystretch{2.2}
\begin{array}{c c}
E_{AI}=\dfrac{\sqrt{2}}{2}E_{DIr};&E_{BI}=\dfrac{\sqrt{2}i}{2}E_{DIr}\\%
E_{AII}=\dfrac{\sqrt{2}i}{2}E_{DII};& E_{BII}=\dfrac{\sqrt{2}}{2}E_{DII}
\end{array}
\end{equation}
Also the beam splitter will recombine the beams into paths $A$ and $B$.
\begin{equation}
\begin{array}{c}
E_{A}=E_{AI}+E_{AII}\\%
E_{B}=E_{BI}+E_{BII}\\%
\end{array}
\end{equation}
Finally, the beams are focused on the detectors.
\begin{equation}
\begin{array}{c}
D_{A}=FT\left(E_{A}\right)\\%
D_{B}=FT\left(E_{B}\right)\\%
\end{array}
\end{equation}
The evolution of the light coming from a central source is pictured in Figure~\ref{fig:I045}.

\newpage
\subsection{Performance definition}
\begin{figure}[h!]%
\begin{center}
\includegraphics[width=0.7\columnwidth]{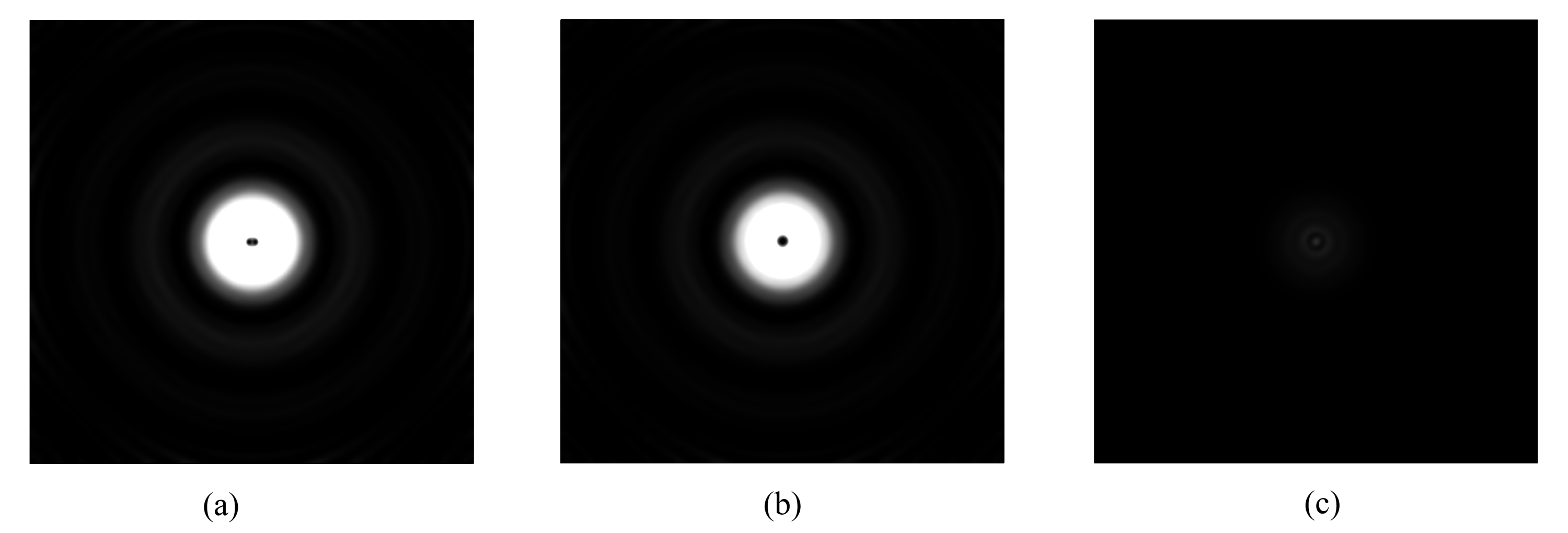}%
\caption{\label{fig:Iglob} Pictures at the detectors for a on-axis source, (a) is for the case of no interferometer ($On_{No}$), (b) is for detector A ($On_{A}$) and (c) is for detector B ($On_{B}$.)}%
\end{center}
\end{figure}

\begin{figure}[h!]%
\begin{center}
\includegraphics[width=0.7\columnwidth]{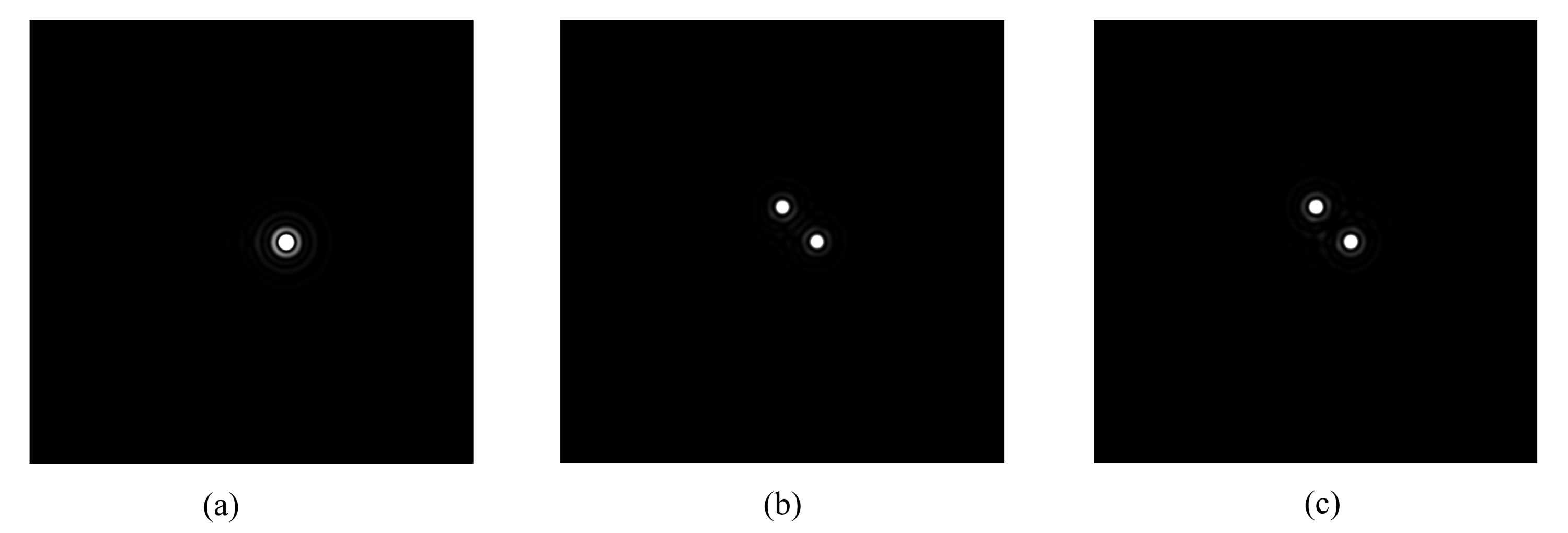}%
\caption{\label{fig:off045} Pictures at the detectors for an off-axis source at  $\alpha=5\lambda/D$, (a) is for the case of no interferometer ($Off_{No}$), (b) is for detector A ($Off_{A}$) and (c) is for detector B ($Off_{B}$.)}%
\end{center}
\end{figure}

\begin{figure}[h!]%
\begin{center}
\includegraphics[width=0.7\columnwidth]{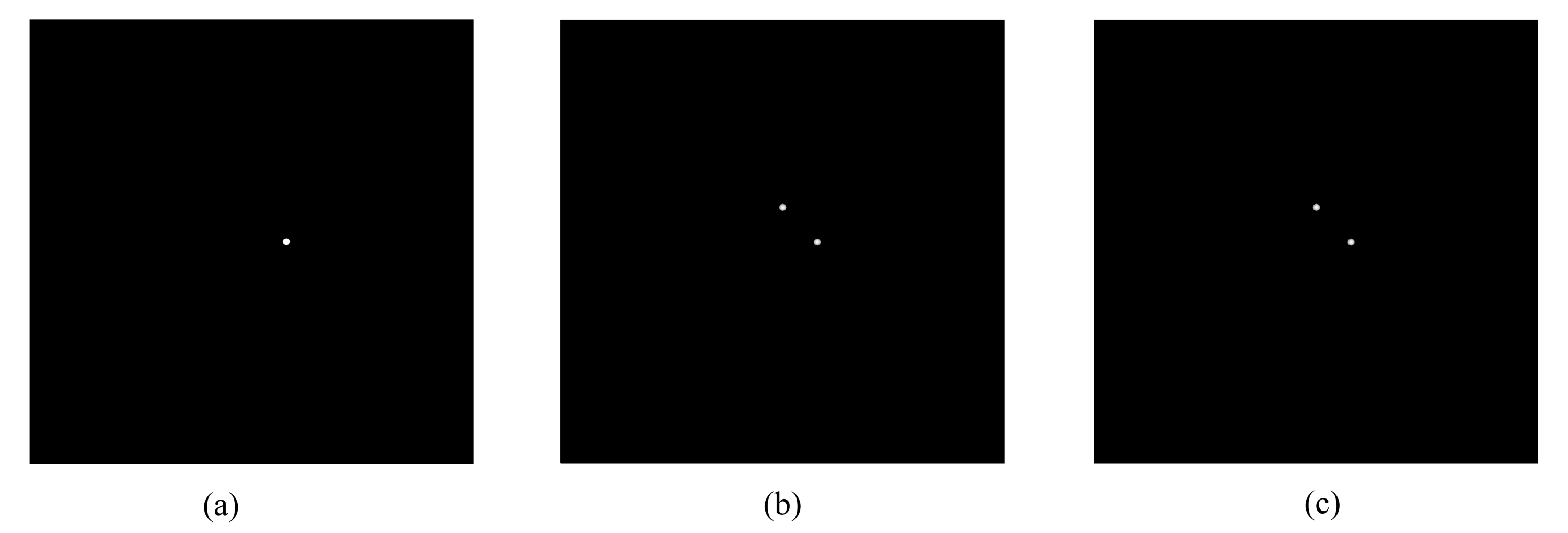}%
\caption{\label{fig:Lobe045} Pictures of the central lobes for an off-axis source at $\alpha=5\lambda/D$, (a) is for the case of no interferometer $Lobe_{No}$, (b) is for detector A and it contains $Lobe_{Ax}$ and $Lobe_{Ay}$ finally,  (c) is for detector B and it contains $Lobe_{Bx}$ and $Lobe_{By}$.}%
\end{center}
\end{figure}

\begin{figure}[h!]%
\begin{center}
\includegraphics[width=0.7\columnwidth]{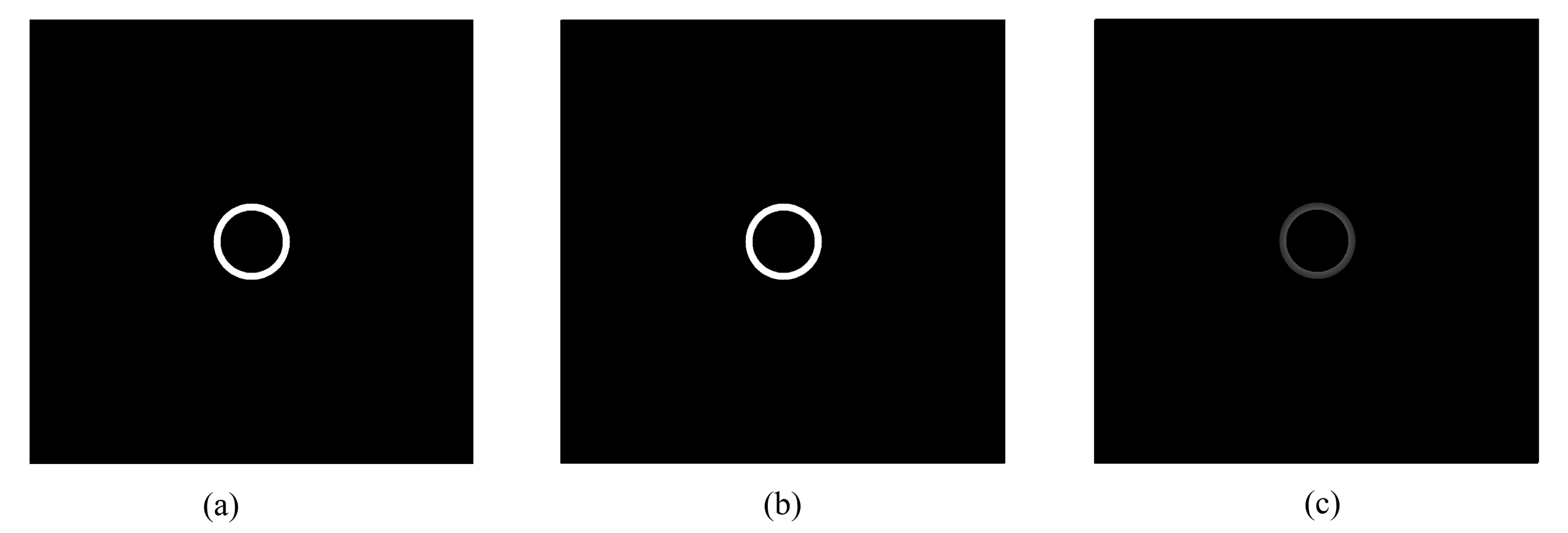}%
\caption{\label{fig:Ring045} Pictures of the rings for the on-axis source for the rings centered on $5\lambda/D$, (a) is for the case of no interferometer ($Ring_{No})$ (b) is for detector A ($Ring_{A}$) and(c) is for detector B  ($Ring_{B}$).}%
\end{center}
\end{figure}

\begin{figure}[h!]%
\begin{center}
\includegraphics[width=0.9\columnwidth]{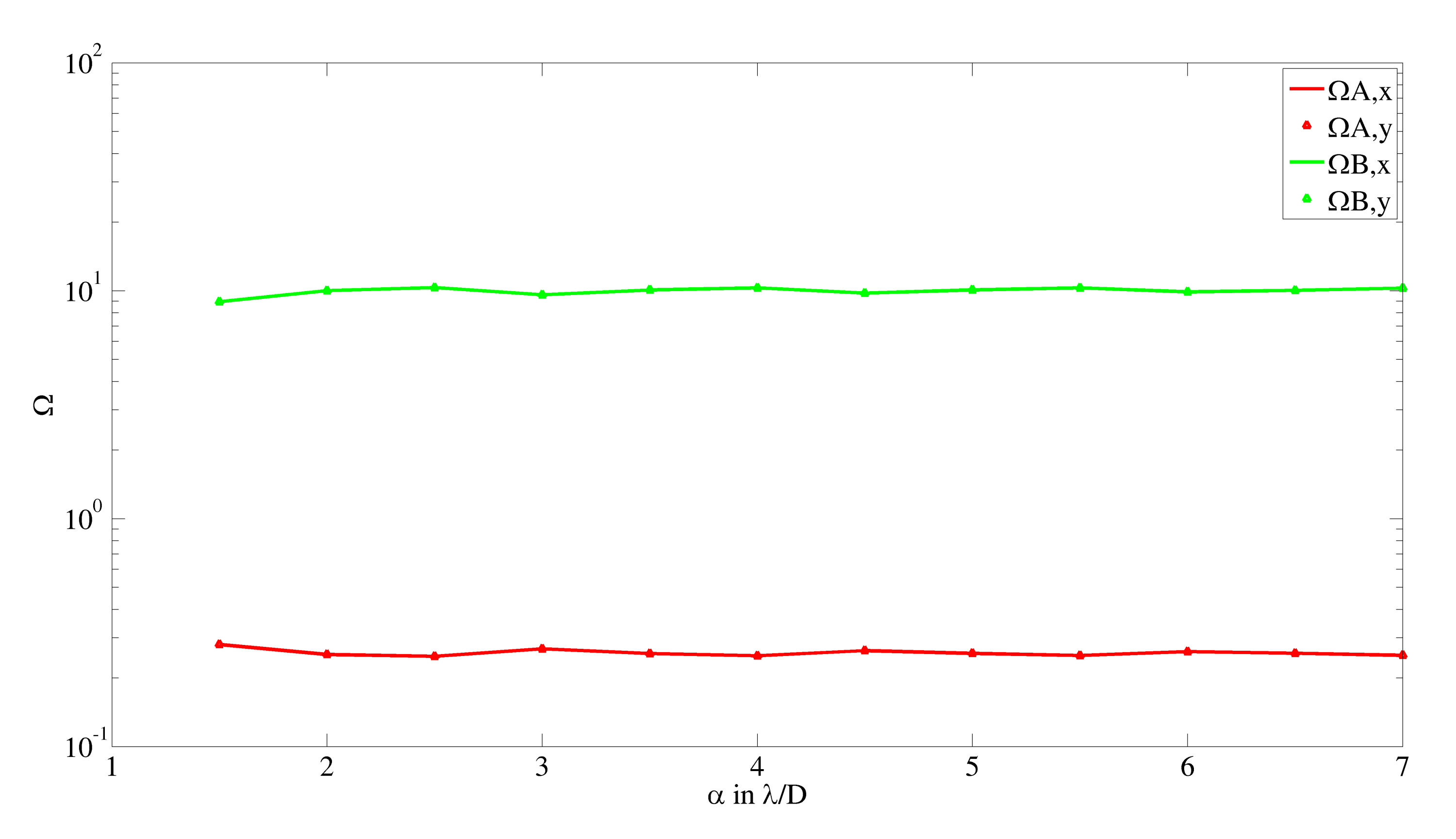}%
\caption{\label{fig:Perf045} Performance of the interferometer for the case of a linearly polarized beam at 45\degree. It can be observed that the performance does not significantly depend on the incidence angle $\alpha$.}%
\end{center}
\end{figure}

To assess the performance of our interferometer, two configurations were simulated: one without the interferometer where the beam after the Lyot stop is directly focused on a detector $D_{0}=FT\left(E_{Lyot}\right)$ the other is the complete setup presented before. Also, we define the performances of the interferometer as a contrast ratio.
The contrast is computed in five steps.
\begin{enumerate}
\item A central source is simulated and its intensity is measured in the three detectors (see Figure~\ref{fig:Iglob})\\%
$\Rightarrow$ $On_{No},\, On_{A},\, On_{B}$ .
\item Off-axis sources are simulated for different incident angles from 1.5 to 7 $\lambda/(D)$. Their intensities are measured on their respective detector (see Figure~\ref{fig:off045}) \\%
$\Rightarrow$ $Off_{No}(\alpha),\, Off_{A}(\alpha), Off_{B}(\alpha)$.
\item The central lobe of each image of the off-axis sources is selected (see Figure~\ref{fig:Lobe045})\\%
$\Rightarrow$ $Lobe_{No},\,Lobe_{Ax},\,Lobe_{Ay},\,Lobe_{Bx},\,Lobe_{By}$.
\item Inside each selected central lobe, the average intensity ($\eta$) is computed.
\[
\begin{array}{c}
\eta_{No}=\left\langle{Lobe_{No}}\right\rangle\\%
\eta_{Ax}=\left\langle{Lobe_{Ax}}\right\rangle;\,\eta_{Ay}=\left\langle{Lobe_{Ay}}\right\rangle\\%
\eta_{Bx}=\left\langle{Lobe_{Bx}}\right\rangle;\,\eta_{By}=\left\langle{Lobe_{By}}\right\rangle\\%
\end{array}
\]
\item On the central sources pictures, rings of a thickness corresponding to the central lobe diameter and centered on the central lobe are selected for each incident angles $\alpha$ (see Figure~\ref{fig:Ring045}).\\%
$\Rightarrow$ $Ring_{No}(\alpha),\, Ring_{A}(\alpha),\,Ring_{B}(\alpha)$.
\item Inside each ring, the average intensity is computed.
\[
\begin{array}{c}
\rho_{No}=\left\langle{Ring_{No}}\right\rangle\\%
\rho_{A}=\left\langle{Ring_{A}}\right\rangle\\%
\rho_{B}=\left\langle{Ring_{B}}\right\rangle
\end{array}
\]
\item The contrast $\Gamma$ is defined as the ratio between the mean intensity contained in the principal lobe of the off-axis source on the mean intensity of the ring produced by the central source containing this central lobe.\\%
\[
\arraycolsep=1.4pt\def\arraystretch{2.2}
\begin{array}{c}
\Gamma_{No}\left(\alpha\right)=\dfrac{\eta_{No}\left(\alpha\right)}{\rho{_No}\left(\alpha\right)}\\%
\Gamma_{Ax}\left(\alpha\right)=\dfrac{\eta_{Ax}\left(\alpha\right)}{\rho_{A}\left(\alpha\right)};\,\Gamma_{Ay}\left(\theta\right)=\dfrac{\eta_{Ay}\left(\alpha\right)}{\rho_{A}\left(\alpha\right)}\\%
\Gamma_{Bx}\left(\alpha\right)=\dfrac{\eta_{Bx}\left(\alpha\right)}{\rho_{B}\left(\alpha\right)};\,\Gamma_{By}\left(\theta\right)=\dfrac{\eta_{By}\left(\alpha\right)}{\rho_{B}\left(\alpha\right)}\\%
\end{array}
\]
\item The performance $\Omega$ is the ratio between the contrast obtained with the interferometer and the contrast obtained without the interferometer for the same incident angle (see Figure~\ref{fig:Perf045}).
\[
\arraycolsep=1.4pt\def\arraystretch{2.2}
\begin{array}{c}
\Omega_{A,x}(\alpha)=\dfrac{\Gamma_{Ax}\left(\alpha\right)}{\Gamma_{No}\left(\alpha\right)};\,\Omega_{A,y}(\lambda)=\dfrac{\Gamma_{Ay}\left(\alpha\right)}{\Gamma_{No}\left(\alpha\right)}\\%
\Omega_{B,x}(\alpha)=\dfrac{\Gamma_{Bx}\left(\alpha\right)}{\Gamma_{No}\left(\alpha\right)};\,\Omega_{B,y}(\lambda)=\dfrac{\Gamma_{By}\left(\alpha\right)}{\Gamma_{No}\left(\alpha\right)}\\%
\end{array}
\]
\end{enumerate}

\subsection{Results}
In the mathematical model, the Jones formalism was used. This formalism is well suited to represent the phase of the beam through the optical setup and the beam transformations. However it only allows to properly represent completely polarized beams, which is not the case for a real observation. To overcome this situation, we decided to test the performance for differently polarized beams to determine whether the incident polarization influences or not the performance. During our simulations, we chose a refractive index $n=1.515$ for the Dove prisms. As seen in Figure~\ref{fig:Perf045}, the performance exhibits a small dependance on the incident angle $\alpha$. Therefore the results are summarized with the mean value of the performance for every incident angle $\alpha$.  The following table presents these results.
\newpage
\begin{table}[h!]
\begin{center}
\caption{\label{tab:polinfl} Results for several incident polarization $P_{in}$ with a 45\degree~difference in the prisms orientation.}
\arraycolsep=1.2pt\def\arraystretch{1.05}
\begin{tabular}{|c|c c c c|}

\hline
&\multicolumn{4}{|c|}{Polarization}\\%
\hline
							&$P_{in}=\mathscr{L}$	&$P_{in}=\mathscr{R}$	&$P_{in}=\leftrightarrow$	&$P_{in}=\updownarrow$	\\%
							\hline
$\Omega_{A,x}$&0.26									&0.26									&0.28											&0.24\\%
$\Omega_{A,y}$&0.26									&0.26									&0.28											&0.26\\%
$\Omega_{B,x}$&9.97									&9.97									&10.87										&9.07\\%
$\Omega_{B,y}$&9.97									&9.97									&9.97											&9.98\\%
\hline
							&$P_{in}=\left(\begin{array}{c}\cos (45)\\\sin(45)\end{array}\right)$&$P_{in}=\left(\begin{array}{c}\cos (-45)\\\sin(-45)\end{array}\right)$&$P_{in}=\left(\begin{array}{c}\cos (30)\\\sin(30)e^{0.5i\pi}\end{array}\right)$&$P_{in}=\left(\begin{array}{c}\cos (30)\\\sin(30)e^{-0.25i\pi}\end{array}\right)$\\%
\hline
$\Omega_{A,x}$&0.26									&0.26									&0.27											&0.27\\%
$\Omega_{A,y}$&0.24									&0.28									&0.26											&0.24\\%
$\Omega_{B,x}$&9.98									&9.97									&10.42										&10.43\\%
$\Omega_{B,y}$&9.07									&9.07									&9.97											&9.42\\%
\hline
&$P_{in}=\left(\begin{array}{c}\cos (30)\\\sin(30)e^{0.3i\pi}\end{array}\right)$&$P_{in}=\left(\begin{array}{c}\cos (70)\\\sin(70)e^{0.65i\pi}\end{array}\right)$&\multicolumn{2} {c|} {}\\%
\hline
$\Omega_{A,x}$&0.27									&0.24									&\multicolumn{2} {c|} {}\\%
$\Omega_{A,y}$&0.25									&0.27									&\multicolumn{2} {c|} {}\\%
$\Omega_{B,x}$&10.43								&9.28									&\multicolumn{2} {c|} {}\\%
$\Omega_{B,y}$&9.51									&10.24								&\multicolumn{2} {c|} {}\\%

\hline
\end{tabular}
\end{center}
\end{table}

\noindent From these results two major statements can be made.
\begin{enumerate}
\item Polarization of the incident beam does not significantly impact the performance.
\item The interferometer performances are relatively poor compared to a perfect nulling case predicted by a $\pi$ phase difference. This lack of performance is due to the change of polarization state and phase retard produced by the Dove prisms, which were taken into account through their Jones matrices\cite{Mor04}.
\end{enumerate}
To have a better understanding of the poor performance, the phase difference of the x and y components after the Dove prisms $\phi^{out}_{x-y}$ was computed and compared to the phase difference after the AGPM $\phi_{x-y}^{in}$ for several incident polarization. This phase variation $\Delta\phi=\phi^{out}_{x-y}-\phi_{x-y}^{in}$ is related to change in the polarization state and the results are presented in Table~\ref{tab:pha}.

\begin{table}[h!]
\begin{center}
\caption{\label{tab:pha} Table of the phase errors $\Delta\phi (rad)$ after the Dove prisms for different incident polarizations and different orientations of the Dove prisms.}%
\begin{tabular}{|c|c c c c|}
\hline
&\multicolumn{4}{|c|}{Polarization}\\%
\hline
$\delta$\degree		&$\mathscr{L}$&$\mathscr{R}$	&$\leftrightarrow$&$\updownarrow$\\%
\hline
0		&$-0.14\pi$		&$-0.14\pi$			&$-0.14\pi$						&$-0.14\pi$	\\%
45	&$-0.0320\pi$	&$0.0320\pi$		&$-0.030\pi$					&$0.0330\pi$\\%
90	&$0.14\pi$		&$0.14\pi$			&$0.14\pi$						&$0.14\pi$\\%
135	&$0.0320\pi$	&$-0.0320\pi$		&$-0.0330\pi$					&$0.0330\pi$\\%
\hline
\end{tabular}
\end{center}
\end{table}

\noindent The constant phase retard observed for $\delta=0$\degre and $\delta=90$\degre led to the improved version of the interferometer presented in the next section.
\newpage
\section{Setup improvement with a customized birefringent plate}
\begin{figure}[h!]
\begin{center}
\includegraphics[width=0.75\columnwidth]{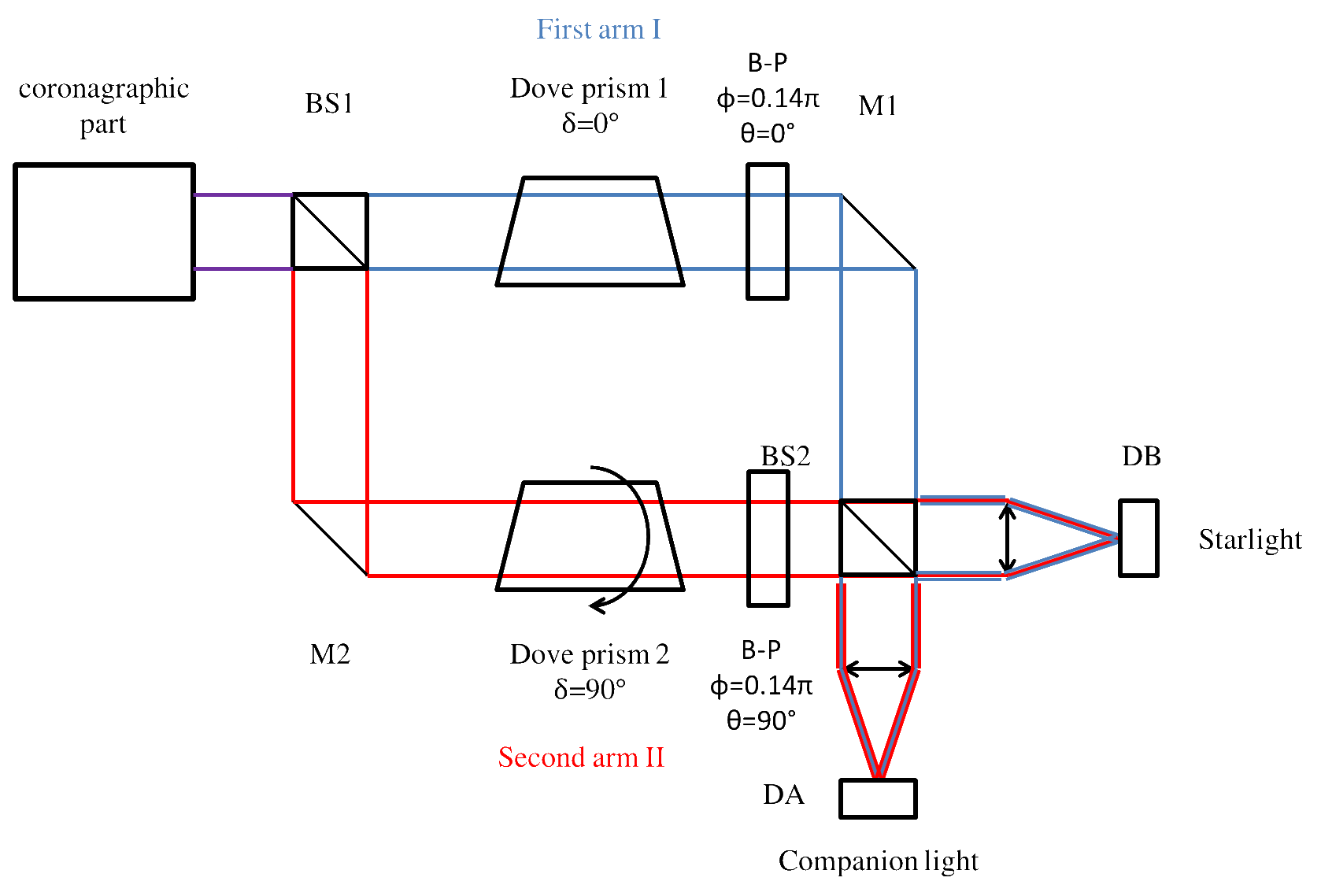}%
\caption{\label{fig:interfpart2} Representation of the modified interferometric setup; the first wave plate is characterized by $\phi=0.14\pi$ and a horizontal fast axis while the other one possesses the same birefringence and a vertical fast axis.}%
\end{center}
\end{figure}

To improve the performance, the phase variation due to the Dove prisms must be reduced. To perform that task we proposed to use a modified version of the setup. The modification consists in a new orientation of the second Dove prism: $\delta_{II}=90$\degree and the addition of two birefringent wave plates to cancel the phase retard $\Delta\phi$ as illustrated in Figure~\ref{fig:interfpart2}. This setup is a modified version of the one proposed by Leach et al.~\cite{Lea02} to sort odd and even OAM. The starlight possessing an OAM of 2 is sent to detector B while the off-axis sources do not possess OAM are sent to detectors A and B.\\%

\begin{figure}[h!]%
\begin{center}
\includegraphics[width=0.9\columnwidth]{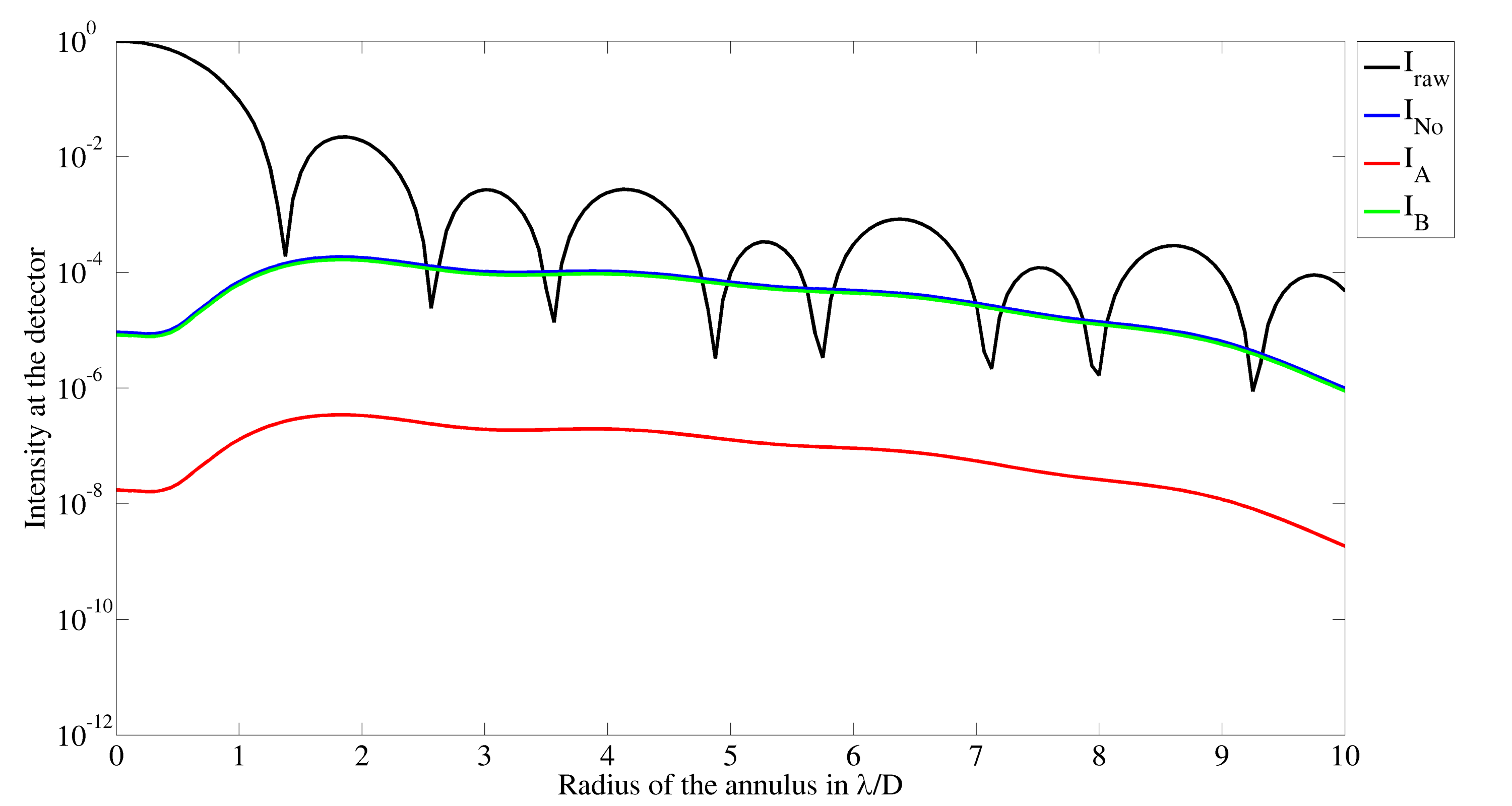}%
\caption{\label{fig:I090} Radial profile of the intensity at the different detectors for a central source; $I_{raw}$ represents the intensity obtained without a coronagraph and without the interferometer. The other intensities are obtained without the interferometric setup ($I_{No}$) and at the two detectors $I_{A},\,I_{B}$. The intensities are normalized with the maximum of $I_{raw}$}%
\end{center}
\end{figure}

\begin{figure}[h!]
\begin{center}
\includegraphics[width=0.5\columnwidth]{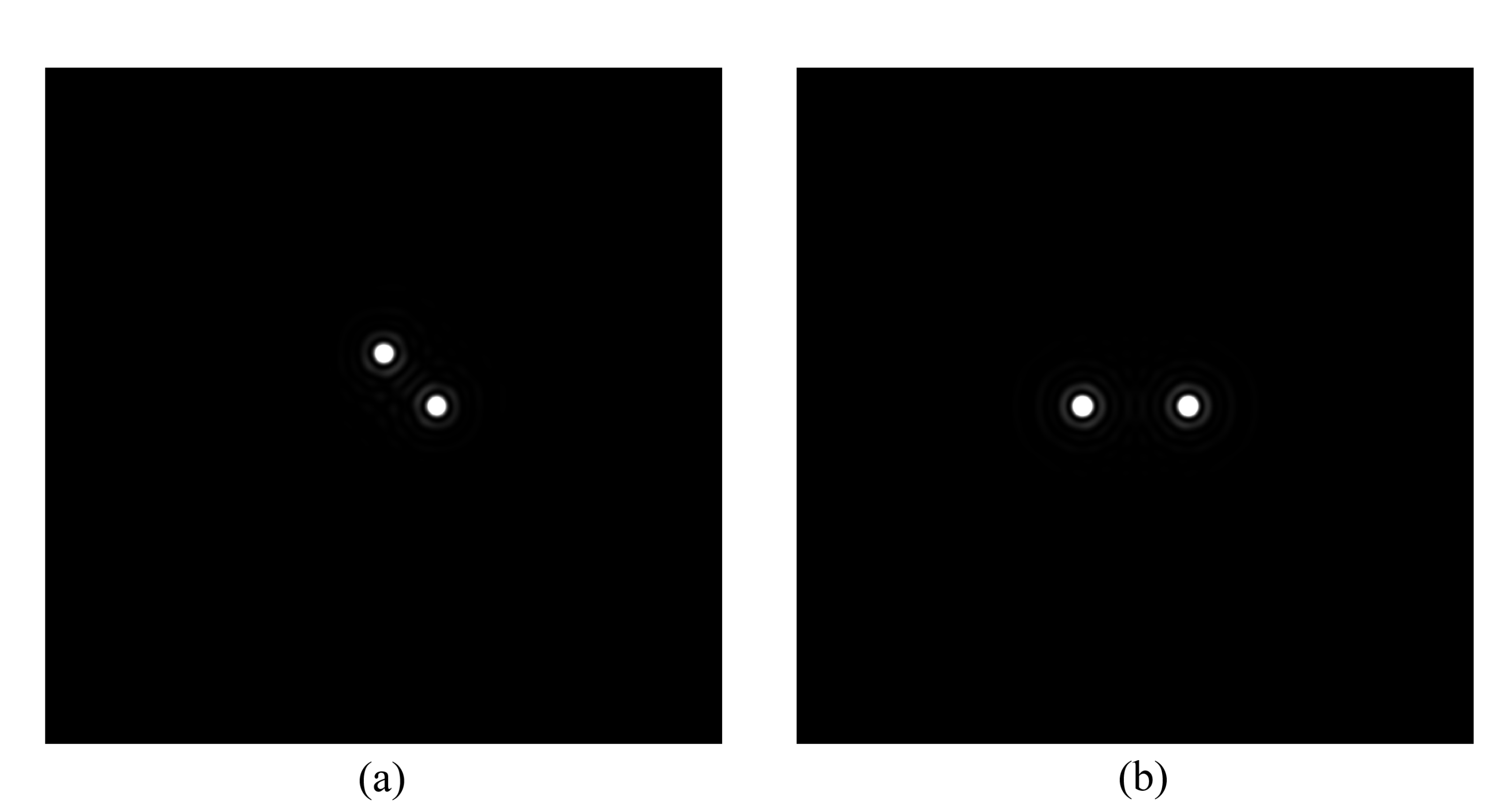}%
\caption{\label{fig:posit} Images at the detector A for the two configurations; (a) stands for the first setup; (b) represents the modified one.}%
\end{center}
\end{figure}

\noindent This time due to the differential orientation of the prisms of 90\degre, the on-axis beams will cancel at detector A and will add at detector B (see Figure~\ref{fig:I090}). Also, the images of the off-axis sources will be on the two sides of the x axis instead of one on the x and the other on the y axis (see Figure~\ref{fig:posit}).
\newpage

\noindent Table~\ref{tab:resinterfmod} presents the results for the modified interferometer for differently polarized incident beams.

\newpage
\begin{table}[h!]
\begin{center}
\caption{\label{tab:resinterfmod} Results for the modified interferometer for several polarized beams $P_{in}$.}
\arraycolsep=1.2pt\def\arraystretch{1.05}
\begin{tabular}{|c|c c c c|}

\hline
&\multicolumn{4}{|c|}{Polarization}\\%
\hline
							&$P_{in}=\mathscr{L}$	&$P_{in}=\mathscr{R}$	&$P_{in}=\leftrightarrow$	&$P_{in}=\updownarrow$	\\%
							\hline
$\Omega_{A,x>0}$&123.39								&124.11								&135.32										&135.32\\%
$\Omega_{A,x<0}$&123.39								&124.11								&113.48										&113.48\\%
$\Omega_{B,x>0}$&0.24									&0.24									&0.27											&0.27\\%
$\Omega_{B,x<0}$&0.25									&0.24									&0.22											&0.22\\%
\hline
							&$P_{in}=\left(\begin{array}{c}\cos (45)\\\sin(45)\end{array}\right)$&$P_{in}=\left(\begin{array}{c}\cos (-45)\\\sin(-45)\end{array}\right)$&$P_{in}=\left(\begin{array}{c}\cos (30)\\\sin(30)e^{0.5i\pi}\end{array}\right)$&$P_{in}=\left(\begin{array}{c}\cos (30)\\\sin(30)e^{-0.25i\pi}\end{array}\right)$\\%
\hline
$\Omega_{A,x>0}$&124.11								&124.11									&128.89										&125.33\\%
$\Omega_{A,x<0}$&124.11								&124.11									&117.98										&121.54\\%
$\Omega_{B,x>0}$&0.24									&0.24										&0.26											&0.25\\%
$\Omega_{B,x<0}$&0.24									&0.24										&0.23											&0.24\\%
\hline
&$P_{in}=\left(\begin{array}{c}\cos (30)\\\sin(30)e^{0.3i\pi}\end{array}\right)$&$P_{in}=\left(\begin{array}{c}\cos (70)\\\sin(70)e^{0.65i\pi}\end{array}\right)$&\multicolumn{2} {c|} {}\\%
\hline
$\Omega_{A,x>0}$&129.04								&131.95&\multicolumn{2} {c|} {}\\%
$\Omega_{A,x<0}$&118.12								&115.21&\multicolumn{2} {c|} {}\\%
$\Omega_{B,x>0}$&0.26									&0.23&\multicolumn{2} {c|} {}\\%
$\Omega_{B,x<0}$&0.23									&0.26&\multicolumn{2} {c|} {}\\%
\hline
\end{tabular}
\end{center}
\end{table}

\noindent From these results it can be observed that the added wave-plates and the new prisms orientation improve the performances by a factor higher than ten. Also, variation due to different polarizations are still present, but the performances are still at the same order near $120\pm 10$. These encouraging results validate the modified version of the interferometric setup.\\%

\section{Conclusions and perspectives}
In this paper we presented the concept of an interferometric setup used to improve the performances of a coronagraphic phase mask. The interferometer uses rotating prisms to create a $\pi$ phase shift between the two arms of the interferometer for a central starlight while the off sources are not superimposed.\\%
Thanks to the destructive interference of the central light, the contrast ratio between faint companions and residual starlight is improved increasing the detection capacity.\\%
The first system only increases the performances by a factor of ten. This factor could be improved to a factor one hundred and more by adding customized birefringent plates and changing the orientation of one Dove prism.\\%
In the future, several improvements will be investigated. The first one concerns the refractive index of the prism, since it impacts the phase error $\Delta\phi$, a first optimization will be to select the material with the smallest $\Delta \phi$ while being transmittive in the useful wavelength domain and inducing a realistic size for the internal reflection. Next is to study the prisms birefringence dispersion to compute $\Delta \phi (\lambda)$ and to deduce how the birefringence of the customized wave plates can accommodate this variation. Another interesting prospect will be to compute the sensitivity of the improved interferometer to the central obstruction ratio and the sensitivity to imperfect coronagraphs.
\appendix
\section{Transmissive coefficients}
\label{app:tran}
\begin{equation}\begin{array}{c}
T_{//}=\left\{\dfrac{\left(4n^{2}\sin\alpha\right)\left(n^{2}-\cos^{2}\alpha\right)^{1/2}}{\left[n^{2}\sin\alpha+\left(n^{2}-\cos^{2}\right)^{1/2}\right]^{2}}\right\} \times \left\{\dfrac{\cos\left(\alpha+\alpha'\right)+i n\left[n^{2}\sin^{2}\left(\alpha+\alpha'\right)-1\right]^{1/2}}{\cos\left(\alpha+\alpha'\right)-i n\left[n^{2}\sin^{2}\left(\alpha+\alpha'\right)-1\right]^{1/2}}\right\}\\%
T_{\perp}=\left\{\dfrac{\left(4\sin\alpha\right)\left(n^{2}-\cos^{2}\alpha\right)^{1/2}}{\left[\sin\alpha+\left(n^{2}-\cos^{2}\alpha\right)^{1/2}\right]^{2}}\right\} \times \left\{\dfrac{n\cos\left(\alpha+\alpha'\right)-i\left[n^{2}\sin^{2}\left(\alpha+\alpha'\right)-1\right]^{1/2}}{n\cos\left(\alpha+\alpha'\right)+i\left[n^{2}\sin^{2}\left(\alpha+\alpha'\right)-1\right]^{1/2}}\right\}
\end{array}
\end{equation}
$T_{//}$ and $T_{\perp}$ depend on the angle of the prism edges $\alpha$, on the refractive index $n$ and on the incident angle for the internal reflection inside the prism $\alpha'=\arcsin\left(\dfrac{\cos\alpha}{n}\right)$.\\%

\section*{Acknowledgments}
This work and the author are funded thanks to a European Research Council funding under the European Union's Seventh Framework Program (ERC Grant Agreement n°337569) .\\%
\bibliography{refrepport}   
\bibliographystyle{spiebib14}   

\end{document}